\documentclass[final,3p,times,twocolumn]{elsarticle}

\usepackage{amssymb}
\usepackage{amsmath, amsfonts, array, mathtools}
\usepackage{mathrsfs}
\usepackage{xcolor}
\usepackage{graphicx}
\usepackage{caption}
\usepackage{subcaption}
\usepackage{physics}
\usepackage{slashed}
\usepackage{braket}
\usepackage{hyperref}
\hypersetup{
  colorlinks   = true, 
  urlcolor     = blue, 
  linkcolor    = blue, 
  citecolor   = blue 
}

\journal{Physical Letters A}

\begin{document}

\begin{frontmatter}



\title{Non-classical correlations between a quantum probe and complex quantum systems in presence of noise}

\author[iiser]{Bijoy John Mathew}
\author[iqc,wloo]{Sanchit Srivastava}
\author[iiser]{Anil Shaji}
\affiliation[iiser]{organization={School of Physics, Indian Institute of Science Education and Research Thiruvananthapuram},
             addressline={Maruthamala P.O., Vithura},
             city={Thiruvananthapuram},
             postcode={695551},
             state={Kerala},
             country={India}}

\affiliation[iqc]{organization={Institute for Quantum Computing},
             addressline={University of Waterloo},
             city={Waterloo},
             postcode={N2L 3G1},
             state={Ontario},
             country={Canada}}
\affiliation[wloo]{organization={Department of Physics and Astronomy},
             addressline={University of Waterloo},
             city={Waterloo},
             postcode={N2L 3G1},
             state={Ontario},
             country={Canada}}



\begin{abstract}
Non-classical correlations generated within a quantum probe system when it interacts with a large, macroscopic system can signal the presence of quantum features in the latter. Theoretical models have considered how entanglement generated in photosynthetic bacteria can be probed using light that interacts with them. More recently, a tardigrade was entangled to a transmon qubit. We consider a detailed model including noise for such systems wherein a small quantum probe interacts with a large system in order to delineate the regimes with respect to coupling strengths and noise levels in which such signatures of quantumness in macroscopic systems can realistically be detected.

\end{abstract}







\end{frontmatter}


\section{Introduction}
\label{intro}
The interaction of a quantum system with another quantum system is described by a joint unitary transformation that typically does not factorise into independent unitaries on each. The joint evolution leads to entanglement or other non-classical correlations developing between the two quantum systems. If one of the two systems is treated as classical, this scenario cannot arise. Quantum information, which has the unique feature that it can lie delocalized across multiple subsystems, cannot be shared between a quantum system and a classical one. This fundamental difference between the interaction between a pair of quantum systems and that of between a quantum system and a classical one can be leveraged to probe the ``quantumness" of one of the systems by observing the other. 

The double slit interference experiment has been the most common approach for exploring the quantum features of systems of increasing size and mass like large molecules~\cite{Clauser1997,Brezger2002,Brezger2003,Hackermuller2003,Arndt2005,Cronin2009,Gerlich2011,Juffmann2012,Eibenberger2013,Silva2016,Kienzler2016}. Probing the large system with a small quantum system provides an alternate approach to exploring the extent to which quantum features are manifest in the large one. Of particular interest is the presence of quantum correlations and other quantum features in complex biological systems, which has long been an area of interest~\cite{Bohr1933,Schrodinger1943,Davies2004}. However, Performing double-slit experiments to probe quantum features even in single-celled organisms, which are highly complex macro-molecular systems, is extremely difficult.

A more practical way of showing the presence of such quantum features is to generate quantum correlations between the large system and a smaller, manifestly quantum system. The presence of these correlations can then be detected in the smaller probe system and treated as a signature of non-classical behavior in the larger one. In~\cite{Krisnanda2018}, the non-classical correlations that may be generated between photons in a beam of light that interacts with photo-responsive bacteria are considered. The idea here is that without manifest quantum behavior, the bacteria cannot mediate the generation of non-classical correlations between a pair of photons, which otherwise would not interact with each other directly. Here, a pair of quantum systems that do not necessarily interact directly becomes the probe of the quantum features of the large system. Such features, if any, are revealed by the generation of non-classical correlations between the two parts of the probe. 

If strong correlations like entanglement are detected in the probe, this would indicate that the larger system is very much quantum in nature. One does not expect such behavior in all cases, but the presence of weaker forms of quantum correlations, like quantum discord~\cite{Henderson2001,Ollivier2001,Modi2012,Adesso2016}, would be sufficient for indicating quantum features in the larger system since discord and other non-classical correlations signify the presence of a quantum superposition involving the probe systems and the large one. This way of probing the larger quantum systems also has the advantage that they typically induce weak or non-demolition measurements on the larger system through a dispersive interaction and do not significantly disturb the state of the larger system.

The authors of~\cite{Krisnanda2018} proposed a thought experiment in which light modes that do not interact directly with each other interact with the bacteria inside a Fabry-Perot cavity. Without explicitly modeling the bacteria, they observe that when steady state is reached, non-zero entanglement between light modes is invariably accompanied by entanglement between the light modes and the bacteria. This was motivated by the framework of~\cite{Krisnanda2017} where it was demonstrated that two systems, $A$ and $B$, coupled to a third system $C$ that functions as a mediator such that the Hamiltonian of the system is of the form $H_{AC} + H_{BC}$, can become entangled to each other only when discord $D_{AB|C}$ is generated during the evolution of the systems.

More recently, in~\cite{Lee2022} experimental detection of entanglement between a tardigrade, a species of organism well renowned for its capacity to survive in harsh settings by entering the cryptobiosis state of life~\cite{Neuman2006,Jonsson2008,Mobjerg2011,Mobjerg2018,Mobjerg2021,Keilin1959,Clegg2001}, to a superconducting transmon qubit is demonstrated. The dynamics of this system can also be viewed in the framework of~\cite{Krisnanda2017} where the qubit and the tardigrade act as the systems $A$ and $B$ that do not interact with each other, and the cavity light mode acts as the mediator system $C$. The entanglement between the cavity mode and the qubit then acts as the witness to entanglement between the qubit and the tardigrade.

In this paper, we model the coupling of a macroscopic system to a quantum probe system in the presence of noise. The focus is on how much of the signatures of quantum features of the macroscopic system imprinted on the probe survive when decoherence effects are relevant. This is, of course, an important consideration in the design of any experiment along these lines. We use the framework of Markovian open dynamics and numerically compute the open evolution. \textcolor{black}{Description of the macroscopic system and its quantum probe using  Gorini-Kossakowski-Sudarshan-Lindblad (GKSL) type master equations~\cite{Breuer2006} is one of the objectives of this work because it allows for incorporating various kinds of noise into our analysis in a straightforward manner. In the following, we consider decay of both the macroscopic system and the probe, as well as dephasing of these systems as representative cases. The models and corresponding equations we consider are, however, capable for incorporating any other kind of noise that may appear in an experimental setup by incorporating appropriate Lindlblad operators. The effect of certain types of noise has been considered in the light-bacteria system already in~\cite{Krisnanda2018}, and we verify that the behavior we obtain using the GKSL master equation is in agreement with that in~\cite{Krisnanda2018}. We also extend the analysis of the effect of noise to the qubit-Tardigrade system. } A full simulation of the macroscopic system, including all of its complexity, is beyond the scope of what can be realistically done. The bacteria or tardigrade have to be, therefore, considered as simplified quantum systems with two or more levels. However, the quantum probe and the noise on it can be modeled in detail. We consider the Light-bacteria system in the next section, followed by the tardigrade-qubit system in section \ref{tardsys}. A brief discussion and our conclusions are in section \ref{conclusion}.

\section{Light-bacteria system}
\label{bactsys}
The framework for the light-bacteria model considered by~\cite{Krisnanda2018}, which we use as a basis for our discussions, is as follows. A species of photosynthetic bacteria is placed within a single-sided, driven Fabry-Perot cavity. Two sets of light modes that normally do not interact with each other are considered. This can be accomplished by choosing light modes with mutually orthogonal polarisations or selecting different frequencies and grouping them into two. These two sets of modes make up systems labeled as $A$ and $B$, whereas the bacteria, operating as mediators that introduce an effective coupling between $A$ and $B$, constitute system $C$. We take the environment affecting the three systems as local and independent. Given that the environment of the light modes is effectively the entire electromagnetic field outside the cavity, the decay of the light modes must be taken into account in addition to the effect of the local environment on system $C$. Considering the environments of the three systems to be independent ensures that they do not become entangled purely as a result of their interactions with their respective environments.

The pigment molecules in the bacteria's chlorosomes that act as excitons and couple to light~\cite{Coles2014,Coles2017} exhibit two distinct peaks at wavelengths $\lambda_I = 750$ nm and $\lambda_{II} = 460$ nm in its extinction spectra in water. The bacteria's light-sensitive region is therefore modeled as two collections (corresponding to the two peaks) of $N$ two-level atoms with transition frequencies of $\Omega_I = 2.5 \times 10^{15}$  Hz and $\Omega_{II} = 4.1 \times 10^{15}$ Hz. These two-level atoms are coupled to each light mode through dipole-like interactions. When $N \gg 1$, such a collection of $N$ two-level atoms can be approximated to a spin-$N/2$ system. These states can be mapped to an effective harmonic oscillator using the Holstein-Primakoff transformation~\cite{Holstein1940}. The Hamiltonian for this system is as follows:

\begin{multline}
H = \sum_m \hbar \omega_m a_m^\dagger a_m + \sum_n \hbar \Omega_n b_n^\dagger b_n \\
+ \sum_{m,n} \hbar G_{mn} ( a_m + a_m^\dagger)( b_n + b_n^\dagger) \label{eqn:bact_ham}
\end{multline}

where $a_m^\dagger$ and $a_m$ are the creation and annihilation operators for the light modes and $b_n^\dagger$ and $b_n$ are the bosonic creation and annihilation operators for the bacteria states. Here $m=1,2,...M$ is the label for the cavity modes, and $n = I,II$ is the label for the two bacteria modes. The third term in the Hamiltonian describes the interaction of the light modes with the bacteria states, with $G_{mn}$ giving the coupling strength of the $m^{\rm th}$ light mode with the $n^{\rm th}$  state of the bacteria. We take the frequency of the $m^{\rm th}$ light mode as $\omega_m = 1.37m \times 10^{15}$ Hz and the coupling strength $G_{mn} = mG_n$ where $G_{I} = 0.04 \times 10^{15}$ Hz and $G_{II} = 0.2 \times 10^{15}$ Hz. In actual experiments, the bacteria have to be suspended in a medium, which can be treated as a standard heat bath. Since bacteria are living organisms, this heat bath is quasi-thermal. Markovian open system dynamics have been assumed for the local environment of the light-sensitive part of the bacteria.

Since there is no entanglement present in the beginning, the initial state of the system is a pure product state of the form:
\begin{equation}
    \ket{\Psi} = \Big(\bigotimes_m \ket{\psi_m}\Big) \otimes \Big(\bigotimes_n \ket{\phi_n}\Big),
    \label{eqn:bact_init}
\end{equation}
where $|\psi_m\rangle$ are states in $n$-dimensional Hilbert spaces corresponding to the four light modes, and $|\phi_n\rangle$ are states in the two-dimensional Hilbert spaces of the two bacteria modes.

The evolution of a density matrix under the Markov approximation is described by the GKSL master equation:
\begin{multline}
    \dot{\rho} = \mathcal{L}(\rho) \\ 
    = i[H,\rho] + \sum_{a>0}\left( L_a^\dagger \rho L_a - \frac{1}{2}L_a^\dagger L_a \rho - \frac{1}{2}\rho L_a^\dagger L_a \right). \label{eqn:mastereqn}
\end{multline}
To identify the relevant Lindblad jump operators, $L_a$, we assume that each light mode decays to the vacuum state by successive emission of photons, with no re-absorption. This decay process is desribed by Lindblad jump operator is $L_m = \sqrt{2\kappa_m}a_m$, where $\kappa_m$ are the decay rates of the cavity modes. Under the same conditions, the jump operator for the bacteria mode is $L_n = \sqrt{2\gamma_n}b_n$ with $\gamma_n$ being the decay rates for bacteria. We note here that this form of jump operators is consistent with the condition that the environment for all the light and bacteria modes are local. The jump operators for each mode couples to independent baths, and the noise effects are summed together since, in general, any kind of cross-coupling during the decay process typically does not happen outside nonlinear regimes. Hence, a jump operator of the form $a_1 \otimes a_2 ... b_I \otimes b_{II}$ which also implies individual modes by correlated losses of single photons each, is not considered because it also allows one mode’s environment to affect another mode.

\textcolor{black}{In general, the GKSL master equations can consider a wide variety of noise relevant to a particular experimental environment by incorporating it using appropriate Lindlblad jump operators, and the analysis is only limited by the computational resources available. To illustrate this, we also consider additional Lindblad jump operators corresponding to ``dephasing" noise of the form $\tilde{L}_m  = \sqrt{2 \tilde{\kappa}_m}a_m a^\dagger_m$ for the light modes and $\tilde{L}_n = \sqrt{2\tilde{\gamma}_n}\sigma_z$ for the two-dimensional bacteria modes.}

\begin{figure}[t]
    \centering
    \includegraphics[width=8cm]{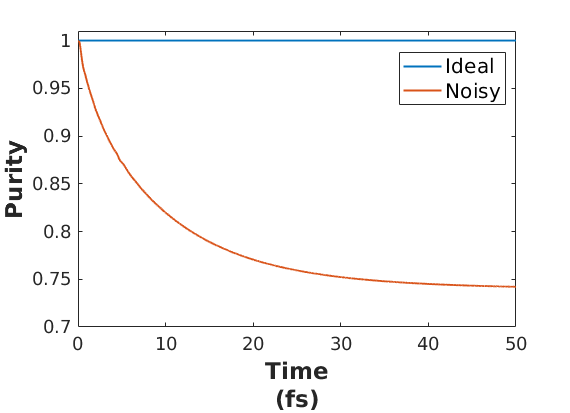}
    \caption{Purity of the total density state with 5-dimensional light modes and 2-dimensional bacteria modes for ideal noiseless evolution (blue) and noisy evolution (red). \textcolor{black}{Both decay as well as dephasing are considered in the latter case.} The plot dips below 1 by an extremely small margin in the ideal case, validating our finite-dimensional model.}
    \label{fig:bact_purity}
\end{figure}

We begin by choosing the ground state for each mode as the initial density matrix of the form in Eq.~\eqref{eqn:bact_init}. The noise level is set by fixing the decay rates at $\kappa_m = \tilde{\kappa}_m = 7.5\times10^{13}$ \textcolor{black}{Hz}, $\gamma_I = \tilde{\gamma}_I = 0.78\times10^{13}$ \textcolor{black}{Hz} and $\gamma_{II} = \tilde{\gamma}_{II} = 3.63\times10^{13}$ \textcolor{black}{Hz}. The evolved state of the system, $\rho(t)$, is obtained by propagating Eq.~\eqref{eqn:mastereqn} numerically with the values of the different parameters involved taken from~\cite{Krisnanda2018}. Various quantities of interest can now be evaluated using the numerically computed density matrix at each time.

\begin{figure}[!htb]
    \centering
    \begin{subfigure}{0.49\textwidth}
        \centering
        \includegraphics[width=0.49\textwidth]{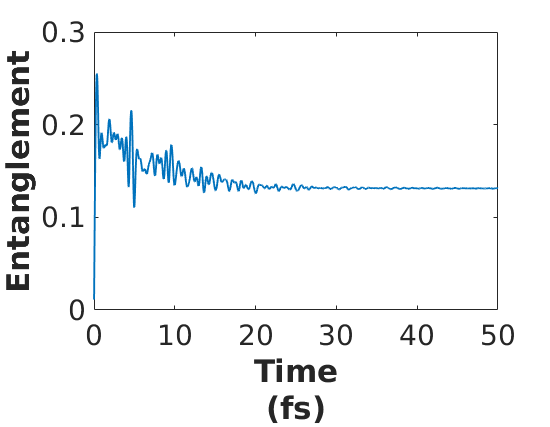}
        \includegraphics[width=0.49\textwidth]{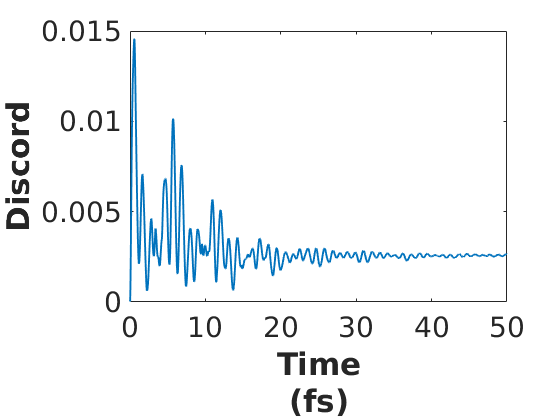}
        \caption{Evolution only under photon emission noise.}
    \end{subfigure}
    \begin{subfigure}{0.49\textwidth}
        \centering
        \includegraphics[width=0.49\textwidth]{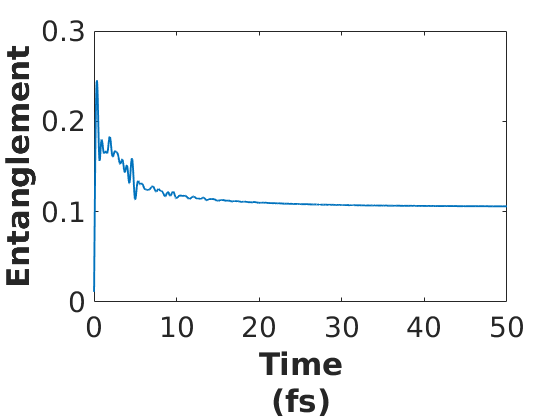}
        \includegraphics[width=0.49\textwidth]{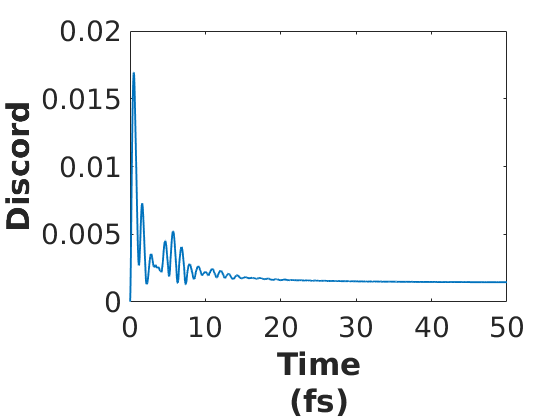}
        \caption{Evolution under both photon emission and dephasing noise.}
    \end{subfigure}
    \caption{\textcolor{black}{The plots on the left show the entanglement between the light modes and bacteria modes, characterized using the negativity while those on the right show the quantum discord evaluated between the two bacteria modes. The top row corresponds to only decay while the bottom row incorporates both decay and dephasing. The values chosen for the relevant parameters for the numerical integration are listed in the main text.}}
    \label{fig:bact_ent}
\end{figure}

We validate our numerics by first confirming that the trace is preserved up to the order of $10^{-18}$. To make computations feasible, we need to consider finite number states for the light modes. This is done by truncating the Hilbert space at the photon number state $\{\ket{n}\}$ to some fixed $n$. \textcolor{black}{An alternate approach would be to work in terms of Gaussian states for the light modes, but since it would make modelling the noise processes more involved, we choose to work with the number states.} Since the bacterium has characteristic transition frequencies, a particular energy scale is associated with the system. This means that the assumption of restricting the light modes to finite dimensions is physically sound, as there cannot be any contribution from higher photon number states beyond a certain energy. We find the optimum number of dimensions, $n$, for the light modes that minimizes the numerical error in calculating the purity of the state in the absence of any decay while at the same time optimizes the computation cost is around 5. We, therefore, fix $n$ to be 5 as an efficient middle ground between high purity and computational difficulty (see Fig.~\ref{fig:bact_purity}). In Fig.~\ref{fig:bact_purity}, we observe that the purity of the density state under the noisy evolution (\textcolor{black}{both decay and dephasing}) of Eq.~\ref{eqn:mastereqn} goes down as expected. Also shown in the figure as a test of the numerical simulation is the purity of the overall bacteria-light state when the evolution is closed. In this case, the purity remains very close to 1 at all times. 

Light-bacteria entanglement is computed using the negativity as a measure~\cite{Vidal2002,Adesso2004}. It should be noted here that when the subsystems between which entanglement is characterized using negativity are larger than qutrits, the absence of negativity does not mean that the subsystems are not entangled~\cite{Horodecki1996,Peres1996}. However, if negativity is non-zero, it does signify the presence of entanglement. We are also able to show non-classical correlations between the two bacteria modes by calculating the discord between them. Since the two-dimensional bacteria modes we consider here are equivalent to qubits, quantum discord can be computed relatively easily~\cite{Luo2008,Akhtarshenas2015}. Time evolution is done up to $50$ fs, considering the computational load. Each run of numerical integration, for a fixed set of parameters, on a 28 core node of a computational cluster took around 60 hours to finish. The light-bacteria entanglement and bacteria-bacteria discord obtained \textcolor{black}{for different types of noise} are shown in Fig.~\ref{fig:bact_ent}.

\begin{figure}
    \centering
    \begin{subfigure}[b]{0.23\textwidth}
        \centering
        \includegraphics[width=\textwidth]{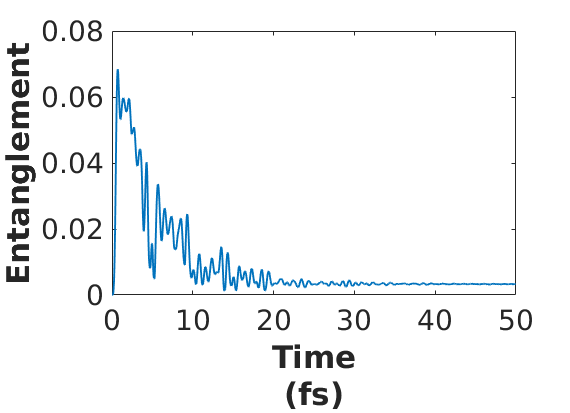}
    \end{subfigure}
    \begin{subfigure}[b]{0.23\textwidth}
        \centering
        \includegraphics[width=\textwidth]{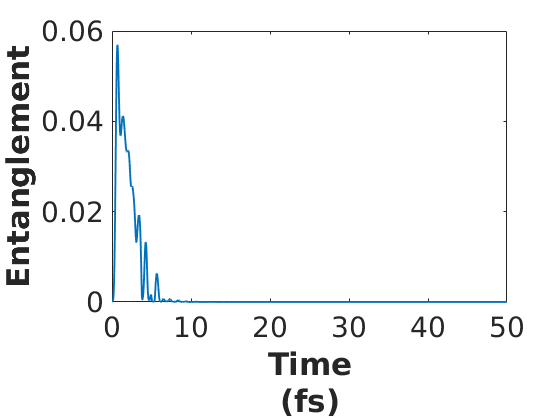}
    \end{subfigure}
    \caption{Entanglement between light modes is shown as a function of time. \textcolor{black}{The plot on the left is with only decay while that on the right is with both decay and dephasing.} Light modes 1 and 2 form one subsystem, while the other comprises of modes 3 and 4. The entanglement between these two subsystems is characterized by the negativity $\mathcal{N}_{12:34}$. Other ways of choosing the two subsystems are also observed to lead to similar results. Due to the size of the subsystems involved, while non-zero values of negativity indicate the presence of entanglement, a zero value does not necessarily mean it is not present.}
    \label{fig:bact_ll_ent}
\end{figure}

We note that the coupled evolution of the bacteria and light does produce quantum entanglement between them. It also produces non-classical correlations between the bacteria modes in our model. The noise tends to destroy both entanglement and discord, however the coupling between the two has the opposite effect of creating such entanglement and non-classical correlations. We note that even in the presence of modest amounts of noise, both entanglement and discord shown in Fig.~\ref{fig:bact_ent} tend to survive and saturate at a constant value. \textcolor{black}{As expected, the saturation values for both entanglement and discord are lower when both decay and dephasing noise are present.} More significant is the light-light entanglement mediated by the bacteria modes. These are observable signatures of the non-classical features in the bacteria modes since these can be detected by measuring only the light coming out of the Fabry-Perot cavity. Fig.~\ref{fig:bact_ll_ent} shows the entanglement between the light modes \textcolor{black}{for different noise models}, indicating that even in the presence of noise, indirect detection of non-classical features in a macroscopic system like bacteria is feasible. 

\section{Qubit-tardigrade system}
\label{tardsys}
The experiment reported in~\cite{Lee2022} consists of placing a tardigrade between the shunt capacitors of a superconducting transmon qubit. The resultant shift in the resonance frequency of the qubit is considered to be due to entanglement generated between the qubit and the tardigrade. In order to quantify the entanglement, this qubit-tardigrade system is entangled with another qubit, and the complete state of the system is reconstructed using quantum state tomography and entanglement measures calculated with the resultant density matrix.

Our analysis focuses on the dynamics of entanglement generation between the tardigrade and the first qubit. This system can be viewed to be analogous to the framework in~\cite{Krisnanda2017}, where the qubit and the tardigrade interaction is mediated by the electromagnetic mode of the microwave cavity in which the system is kept. In this way, entanglement detected in the cavity mode with respect to the qubit-tardigrade partition will imply the generation of non-classical correlations, including entanglement between the qubit and the tardigrade.

The tardigrade is modeled in a manner similar to the bacteria in the previous section, as a collection of dipoles coupled to the cavity field and, hence, is approximated as harmonic oscillators. The Hamiltonian for this system can be written as:
\begin{eqnarray}
    H_{qt} & = &  \frac{\hbar\omega_q}{2}\sigma_z + \sum_l \hbar\omega_l a_l^\dagger a_l + \sum_t \hbar\omega_t b_t^\dagger b_t^{\phantom{\dagger}} \nonumber \\
    && \quad + \, \sum_l \sigma_x \hbar g_{ql} \big(a_l^{\phantom{\dagger}} + a_l^\dagger\big)   \nonumber \\
    && \qquad + \, \sum_t \hbar f_{tl} \big(b_t^{\phantom{\dagger}} + b_t^\dagger \big) \big( a_l^{\phantom{\dagger}} + a_l^\dagger \big) \label{eqn:tard_ham}
\end{eqnarray}
where the first term is for the qubit, the second corresponds to the surrounding electromagnetic field, while the third term describes the dipoles inside the tardigrade. The fourth and fifth terms represent the coupling between the qubit and the electromagnetic field and the coupling between the tardigrade dipoles and the electromagnetic field, respectively. There is no direct interaction between the qubit and the tardigrade. The qubit frequency, $\omega_q$, is taken from~\cite{Lee2022} to be $3.271 \times 10^{9}$ Hz. In order to obtain reasonable dynamics, the light and tardigrade mode frequencies and coupling strengths are taken to be comparable to the qubit frequency. The values are fixed at $\omega_l = 4.521 \times 10^9$ Hz and $\omega_t = 2.7 \times 10^9$ Hz. The coupling strengths are varied within the interval $g_{ql}, f_{tl} = [0,0.3]\times10^{9}$ Hz. 

\begin{figure}[t]
    \centering
    \begin{subfigure}[b]{0.49\textwidth}
        \centering
        \includegraphics[width=0.49\textwidth]{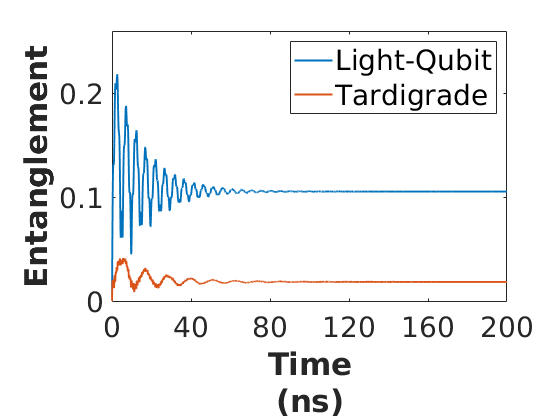}
        \includegraphics[width=0.49\textwidth]{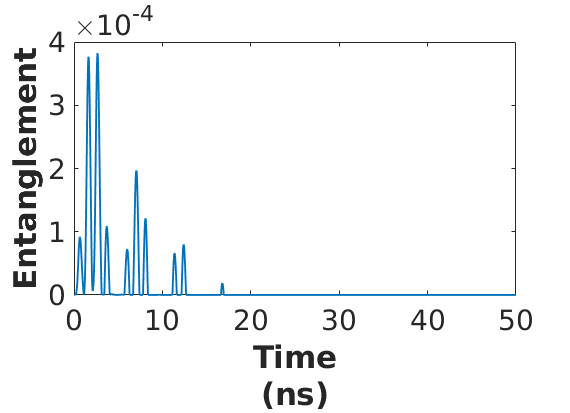}
        \caption{Time evolution of negativity with only photon emission noise.}
        \label{fig:tard_stdnoise_allneg_old}
    \end{subfigure}
    \begin{subfigure}[b]{0.49\textwidth}
        \centering
        \includegraphics[width=0.49\textwidth]{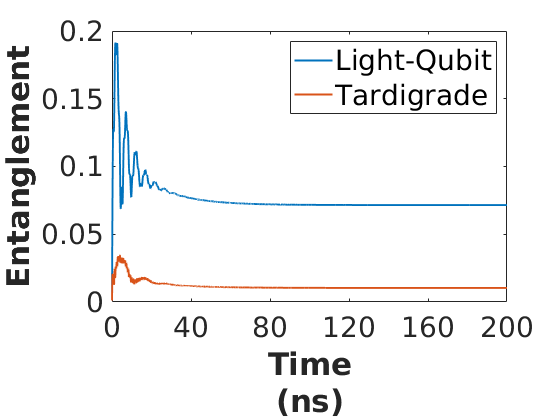}
        \includegraphics[width=0.49\textwidth]{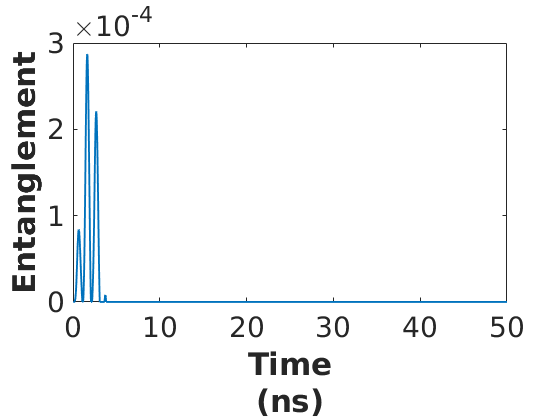}
        \caption{Time evolution of negativity with photon emission and dephasing noise.}
        \label{fig:tard_llneg_old}
    \end{subfigure}
    \caption{\textcolor{black}{The plots on the left show time evolution of the entanglement (negativity) between the light modes and qubit mode (blue), and Tardigrade mode plus the rest of the system (red). The light-qubit and light-Tardigrade coupling constants are fixed. The plots on the right show the time evolution of negativity between the two light modes considered. The top row corresponds to only decay while the bottom row includes both decay and dephasing.}}
    \label{fig:tard_negs}
\end{figure}

For our analysis, we chose only one mode for the tardigrade and two modes for the cavity for computational simplicity. The initial state is written as:
\begin{equation}
    |\Psi\rangle = |\psi_l \rangle \otimes |\phi_t \rangle \otimes |\phi_q\rangle \label{eqn:tard_init}
\end{equation}
where $|\psi_l\rangle$ and $|\phi_t\rangle$ are states in Hilbert spaces of dimension $n_l$ and $n_t$ respectively, describing cavity mode and tardigrade, while $|\phi_q\rangle$ is a state in the two-dimensional Hilbert space of the qubit. Each of these systems is taken to be in their respective ground states initially.

The Lindblad jump operators are identified using similar arguments as in the earlier section. We assume that photons are successively emitted from each of the modes and escape the system without interacting further, and no other photon absorption occurs. Then the jump operators are $L_l = \sqrt{2\kappa_l}a_l$ for light modes, $L_t = \sqrt{2\gamma_t}b_t$ for the tardigrade mode and $L_q = \sqrt{2\delta_q}\sigma_{-}$ for the qubit mode, where $\sigma_{-} = \ketbra{0}{1}$, and $\kappa_l, \delta_q$ and $\gamma_t$ are the respective decay rates. \textcolor{black}{Similarly, the Lindblad jump operators for dephasing noise are $\tilde{L}_l = \sqrt{2\tilde{\kappa}_l}a_l a^{\dagger}_l$ for light modes, $\tilde{L}_t = \sqrt{2\tilde{\gamma}_t}b_t b^{\dagger}_t$ for the Tardigrade mode and $\tilde{L}_q = \sqrt{2\tilde{\delta}_q}\sigma_z$ for the qubit mode.} 

\begin{figure}[t]
    \centering
    \begin{subfigure}{0.45\textwidth}
        \centering
        \includegraphics[width=7cm]{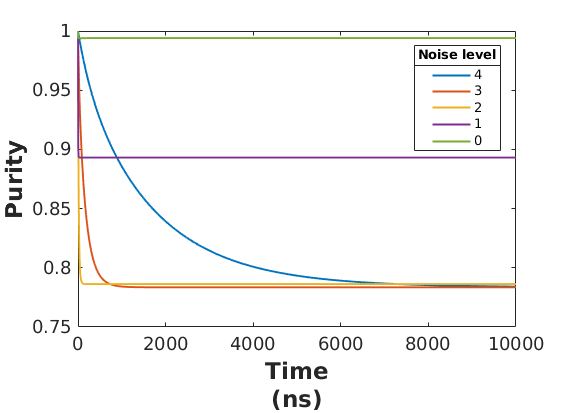}
        \caption{}
        \label{fig:tard_noise_pure}
    \end{subfigure}
    \begin{subfigure}{0.45\textwidth}
        \centering
        \includegraphics[width=7cm]{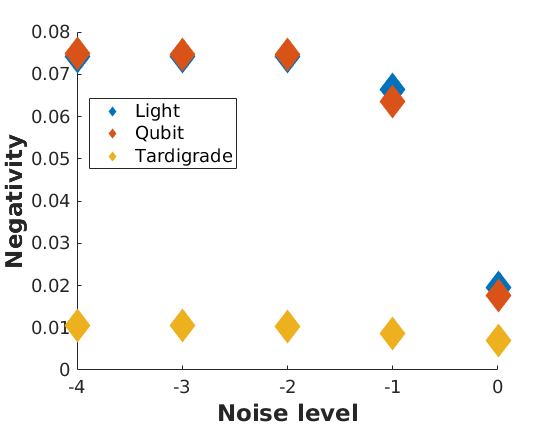}
        \caption{}
        \label{fig:tard_noise_stedvar}
    \end{subfigure}
    \caption{(a) Purity of the state of the qubit-tardigrade system under noisy evolution for different noise levels. (b) Scatter plot of steady state negativity for each subsystem corresponding to different noise levels defined by the decay rates, $[\kappa_l, \gamma_t, \delta_q] \sim [10^{9-i},10^{9-i},10^{9-i}]$. The $x$-axis of the plot is labeled by $-i$, where $i$ determines how many orders of magnitude difference there is between the decay rate and the typical frequency of the tardigrade and qubit of $10^9$ Hz.}
    \label{fig:tard_noise}
\end{figure}

A GKSL master equation in the same form as Eq.~\eqref{eqn:mastereqn} is numerically integrated using the initial state from Eq.~\eqref{eqn:tard_init}. We first verify that the trace of the density matrix for the whole system is preserved within extremely small numerical errors ($\sim10^{-15}$) to validate the accuracy of the numerical integration. The dimension of the Hilbert spaces of cavity light mode and tardigrade mode are both truncated at $n_l=n_t=5$ which is found optimal for enabling the computations within reasonable time while minimizing errors in the purity of the overall state under closed evolution, which only dips below 1 by an extremely small amount during the integration time just as in the light-bacteria system.

\begin{figure*}[t]
    \centering
    \begin{subfigure}{0.99\textwidth}
        \centering
        \includegraphics[width=0.3\textwidth]{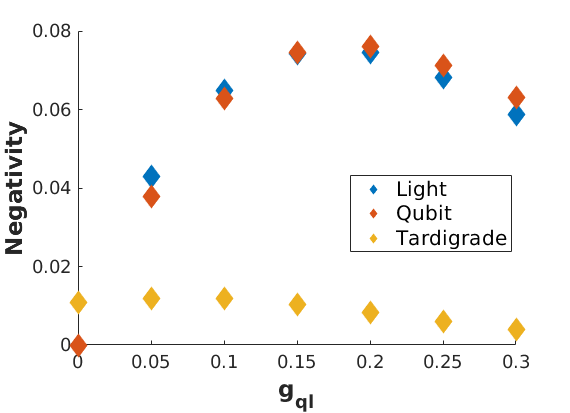}
        \includegraphics[width=0.3\textwidth]{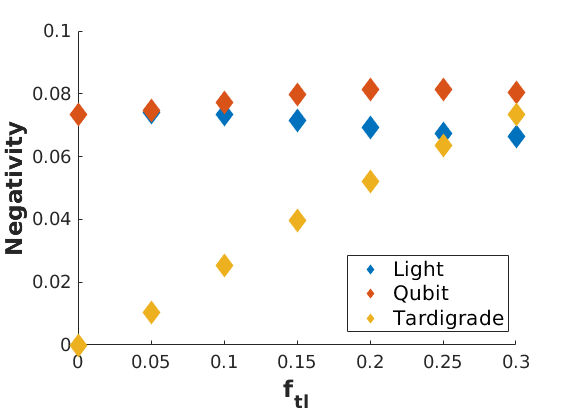}
        \includegraphics[width=0.3\textwidth]{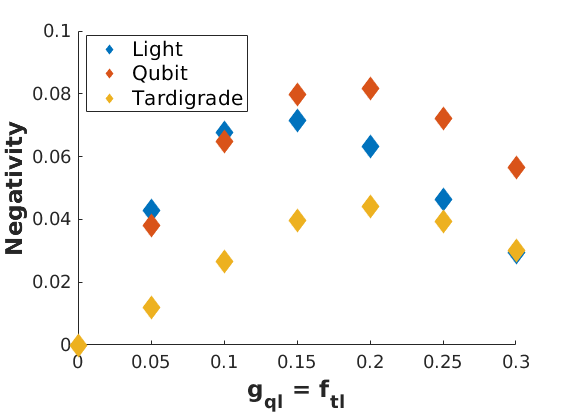}
        \caption{Noise level $\sim 10^7$}
        \label{fig:tard_steadies7}
    \end{subfigure}
    \begin{subfigure}{0.99\textwidth}
        \centering
        \includegraphics[width=0.3\textwidth]{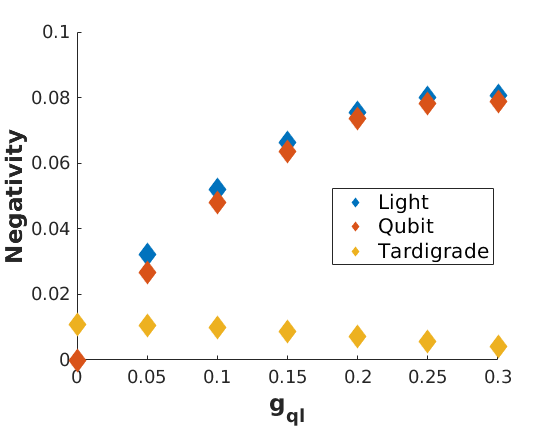}
        \includegraphics[width=0.3\textwidth]{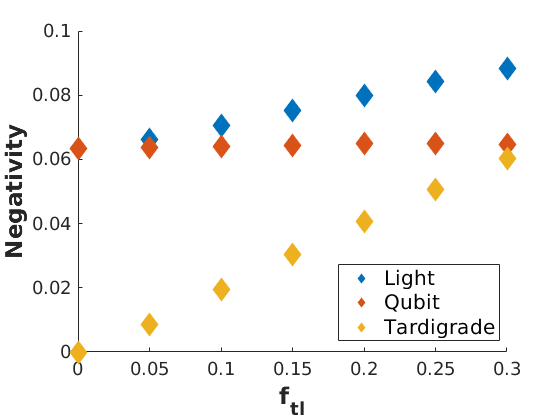}
        \includegraphics[width=0.3\textwidth]{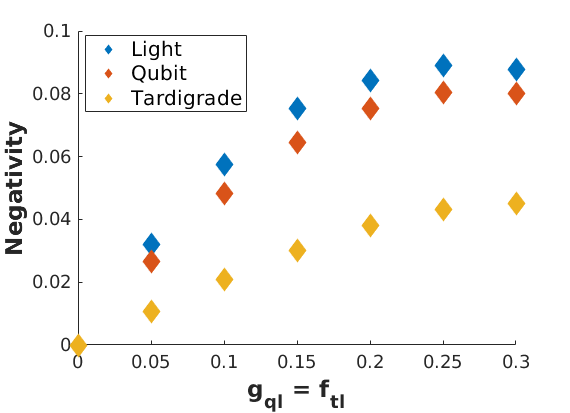}
        \caption{Noise level $\sim 10^8$}
        \label{fig:tard_steadies8}
    \end{subfigure}
    \begin{subfigure}{0.99\textwidth}
        \centering
        \includegraphics[width=0.3\textwidth]{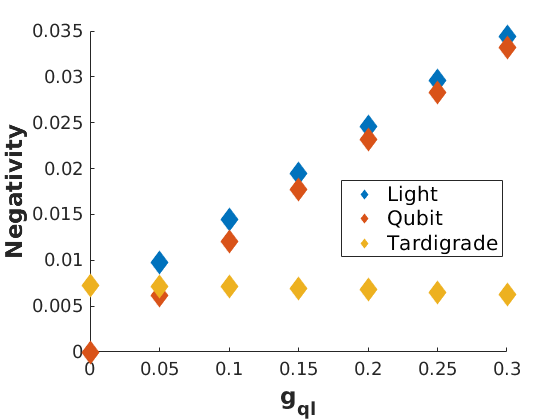}
        \includegraphics[width=0.3\textwidth]{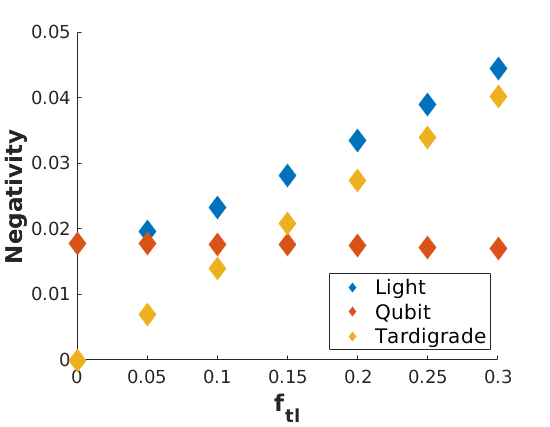}
        \includegraphics[width=0.3\textwidth]{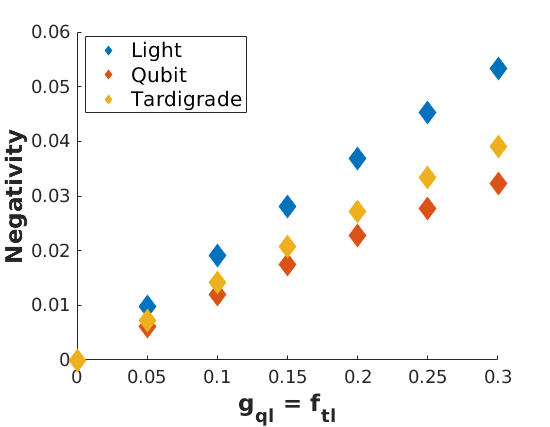}
        \caption{Noise level $\sim 10^9$}
        \label{fig:tard_steadies9}
    \end{subfigure}
    \caption{Scatter plot of steady state values of the negativity for each of light (blue), qubit (red), and tardigrade (yellow) subsystems at different levels of noise in (a), (b), and (c). The plots on the left show the variance with the light-qubit coupling, $g_{ql}$, while the light-tardigrade coupling is kept at $f_{tl} = 0.05$. Middle plots show the variance with the light-tardigrade coupling, $f_{tl}$, while the light-qubit coupling is kept at $g_{ql} = 0.15$. Right plots show the variance with both the couplings, which are kept the same, $f_{tl} = g_{ql}$.}
    \label{fig:tard_steadies}
\end{figure*}

The time evolution plot of the negativity measure of each of the light, qubit, and Tardigrade subsystems \textcolor{black}{in the presence of decay and dephasing} is shown in Fig.~\ref{fig:tard_negs}. Here, the coupling strengths are fixed at $g_{ql} = 0.15 \times 10^{9}$ Hz and $f_{tl} = 0.05 \times 10^{9}$ Hz. The noise level is set by fixing the decay rates at $\kappa_l = \tilde{\kappa}_l = 3.7 \times 10^{7}$ \textcolor{black}{Hz}, $\delta_q = \tilde{\delta}_q = 2.5 \times 10^{7}$ \textcolor{black}{Hz}, and $\gamma_t = \tilde{\gamma}_t = 1.8 \times 10^{7}$ \textcolor{black}{Hz}, respectively. The system is evolved till time, $200ns$, in order to get steady state values for negativity. In this case, each run of code for a fixed set of parameters on a computational node with 28 cores took around 6 hours on average; hence, unlike the case for the previous section, it was more computationally feasible here to do further analysis based on varying coupling strengths and levels of noise, due to the much smaller dimensional structure arising from this model having only one light mode under consideration.

The next analysis is of the steady state values for the negativity of each subsystem, varied for discrete levels of noise (\textcolor{black}{both decay and dephasing}) and coupling strengths. We vary the decay rates in terms of different orders of magnitude below the respective frequencies of each subsystem. Coupling constants are varied in equal intervals within the range of $[0,0.3]\times10^{9}$ \textcolor{black}{Hz}.

Fig.~\ref{fig:tard_noise_pure} plots the time evolution of the purity of the density state for different noise levels, and Fig.~\ref{fig:tard_noise_stedvar} is a scatter plot of the steady state values of the negativity measure of each subsystem varied against different noise levels. The coupling strengths for both figures are fixed as before. It can be seen that, for noise levels up to two orders of magnitude lower than the frequency order, the purity of the system tends to approach 0.78 asymptotically. However, when the decay rates are closer to the order of magnitude of the frequencies, the system loses little of its purity. Similarly, the steady state values of the negativity of each subsystem appear to be constant until the decay rates get closer to the order of the frequencies, when they sharply decrease. 

Finally, in Fig.~\ref{fig:tard_steadies}, three sets of scatter plots show the behavior of the steady state values of negativity for each subsystem with respect to the coupling strengths for different \textcolor{black}{levels of noise (both decay and dephasing)}. In each row, the light-qubit coupling, $g_{ql}$, is varied in the first scatter plot, keeping $f_{tl}$ fixed. In the second plot on each row, the light-tardigrade coupling, $f_{tl}$, is varied, and finally, in the third scatter plot, both couplings are varied while keeping them equal to each other. We observe that the steady state values of the detectable entanglement (particularly that between the light modes) increase with increase in the relevant couplings while they go down with increasing noise as expected. 

\section{Conclusions}
\label{conclusion}

We have examined in detail the interesting idea put forward in~\cite{Krisnanda2018} that non-classical correlations induced in a microscopic quantum probe by its interaction with a macroscopic system can be used to detect the presence of quantum features in the macroscopic system. We focused on a case where the quantum probe is continually interacting with the macroscopic system, and at the same time, both are subject to environmental noise and decoherence. Our objective was to explore the robustness of this way of detecting quantum features in the macroscopic system against noise, which is to be expected in most realistic scenarios. In place of the heuristic models for noise used in previous works along similar lines, we have used a GKSL master equation to describe the open dynamics of the probe and macroscopic system. We numerically integrated the relevant master equations using realistic parameter values and observed that, indeed, entanglement and other non-classical correlations are built up between the quantum probe and various components of the macroscopic system. We also verified that these correlations are robust against noise and saturate to finite values for moderate and low levels of noise in the system. 

Of particular interest is the non-classical correlations generated between the subsystems that constitute the quantum probe, starting from a product state of these subsystems. These are the detectable signatures of quantum features in the macroscopic system because, without such quantum features, non-classical correlations cannot appear between probe subsystems. We have shown that, indeed, such correlations are generated, and even with noise, they retain finite saturation values. We also explored how these saturation values vary when the coupling between various subsystems, as well as the noise levels, are changed. We see that under diverse conditions, the non-classical correlations in the quantum probe remain finite and detectable. This gives further confidence that in realistic experimental scenarios, it may be possible to elucidate signatures of quantum features present in eminently macroscopic systems like the bacteria and tardigrade we have discussed through the detection of such correlations in a quantum probe that interacts with it. 

The detailed analysis with respect to varying coupling strengths and noise levels gives helpful pointers to possible experimental setups that are designed to study non-classical correlations in such complex systems. The model we used is still relatively simple owing to the computational complexity of integrating the master equation. Future avenues of research include optimizing the model, adding more realistic effects, as well as investigating the effects of varying transition frequencies and more modes for each subsystem that presently considered. We have used negativity as a quick and ready measure of entanglement throughout the manuscript, which was sufficient since the negativity was always non-zero in the cases we considered. Due to the large dimensions of the Hilbert spaces involved, a zero value of negativity does not necessarily imply zero entanglement. In such cases, other measures of entanglement that are typically harder to compute may have to be employed. This is also a consideration for the future that is expected to be relevant if higher noise levels are to be explored. 

\section*{Acknowledgements}
This work was supported by the QuEST program of the Department of Science and Technology through project No. Q113 under Theme 4. The authors acknowledge IISER TVM and Centre for High-Performance Computing for the use of the {\em Padmanabha} computational cluster.

\label{conclude}


\bibliographystyle{elsarticle-num} 
\bibliography{biblio}




\end{document}